# AN INTEGRATED METHOD OF DATA HIDING AND COMPRESSION OF MEDICAL IMAGES

M.MaryShanthi Rani and S.Lakshmanan

Department of Computer Science and Applications,
The Gandhigram Rural Institute – Deemed University,
Gandhigram – 624 302,Dindigul, Tamil Nadu, India

*ABSTRACT*

*A new technique for embedding data into an image coupled with compression has been proposed in this paper. A fast and efficient coding algorithms are needed for effective storage and transmission, due to the popularity of telemedicine and the use of digital medical images. Medical images are produced and transferred between hospitals for review by physicians who are geographically apart. Such image data need to be stored for future reference of patients as well. This necessitates compact storage of medical images before being transmitted over Internet. Moreover, as the patient information is also embedded within the medical images, it is very important to maintain the confidentiality of patient data. Hence, this article aims at hiding patient information as well, within the medical image followed by joint compression. The hidden data and the host image are absolutely recoverable from the embedded image without any loss.*

## 1. INTRODUCTION

Secure and fast transmission of medical images is a great challenge in the field of telemedicine. Nowadays, in order to maintain the confidentiality of patient data, they are embedded within the medical image itself. The process of hiding data in an image is called steganography which is derived from the Greek language and means covert writing. It is the technique of encoding secret information in a communication channel in such a manner that the very existence of the information is concealed. Computer-based image steganography is one way of data hiding which provides data security in digital images. The aim is to embed and deliver secret messages in digital images without any suspiciousness. The secret message might be a caption, a plain text, another image, a control signal, or anything that can be represented in bit stream form [2].

Picture Achieving Communication System (PACS) is used for the transmission of medical images. PACS compress the medical images for transfer to other hospitals or storing locally. It saves the storage space of database.

The objective of image compression is to reduce the redundancy of the image and to store or transmit data in an efficient form. The main goal of such system is to reduce the storage quantity as much as possible. As in telemedicine, videos and the medical images are transmitted through advanced telecommunication links, medical image compression without any loss of useful information is of immense importance for the fast transfer of the information [3].

Generally, lossless compression techniques are used for compressing medical images to preserve the original image. Lossless compression is a compression algorithm in which the original image

International Journal of Advanced Information Technology (IJAIT) Vol. 6, No. 1, February 2016

can be completely recovered from the compressed image without any loss. Some of the popular lossless compression techniques are

- Run length coding
- Huffman coding
- Arithmetic encoding
- Entropy coding
- Area coding

Run length coding is a simple method used for compressing sequential data. It achieves compression by eliminating redundancy and avoiding repetitive data.

The combination of a run-length coding scheme followed by a Huffman coder forms the basis of image compression standards. These compression standards yield good compression ratios from 20:1 to 50:1.

In this proposed method Run Length Encoding is used to compress the image. The remaining part of this article is organized as follows, Section 2 discusses about related work in this research area, section 3 describes the proposed method, Section 4 analyses the performance of the proposed method and Section 5 presents the conclusion and future work.

## 2. RELATED WORK

In 2008,M.A.Ansari and R.S.Anand presented various techniques of image compression in Telemedicine [4]. This paper deals with basic redundancies used to achieve compression specifically focussing on medical image.

The secret data [5] is hidden into the quantized DCT co-efficients of the JPEG compressed image. Finally, to produce the JPEG compressed stego-image, the entropy decoding method is employed.

If the data is hidden in LSB as in [6], a hacker can retrieve the hidden message by tracing the LSBs of all pixels and fix them into bytes. To avoid this and to increase the power of security, the stego image with hidden data is encrypted and the encrypted image is hidden in other images at random based on random number generator.

In yet another method, LSB algorithm is represented with pseudo random number generator which is protected from hackers [11].Random number generator is used as a selector in placing the message bits in image randomly. In this proposed method Secret key need not be shared between the sender and receiver to retrieve the data.

## 3. Proposed Method

In this paper, a new method of hiding text data into an image is proposed. This method is highly suited for hiding patient details into a medical image producing a stego image. Furthermore, the stego image is compressed using lossless compression technique for efficient and fast transmission. In this method, lossless compression technique run-length encoding is used for compressing the stego medical image so as to reconstruct the image at receiving end without any loss in image details, whichare very vital for medical diagnosis. The special feature of this algorithm is data hiding combined with compression. It would be useful specifically for PACS. The algorithm of the proposed method is implemented in three phases. Phase one performs the text hiding in medical image, phase two compresses the stego medical image and phase three performs decompression and hidden data retrieval.

The steps of the proposed algorithm is outlined as follows

### 3.1 Phase One – Data Hiding

Steps:

- Read the input medical image.
- Convert the image into gray scale image.
- Take out the Region of Interest by locating the rectangle enclosing it, with its top left and bottom right co-ordinates as shown in Figure.2 (b).
- Read the text data (for example: Patient id and Name)
- Convert each character of text into its consequent ASCII value as given in Table 1.
- Count the ASCII values to avoid replication of hiding same text characters.
- Find the zero valued pixel p(x,y) surrounded by pixels at four corners( Top ,Bottom, Left ,Right) with value 0.
- Hide the ASCII value at p(x,y), resulting in stego image.

### 3.2 Phase Two – Image compression

After the data hiding process in phase one, the stego image is compressed using Run length encoding and transmitted through the internet.
Steps:

- The stego image is converted into a vector Figure 4 (a).
- Using Run Length Encoding, the above vector is converted into two vectors indicating the element (pixel value) and its run length as shown in Figure 4(b) & (c)

The sample result is shown in Figure 5.

### 3.3 Phase Three – Decompression and Data Retrieval

Phase Three decodes the image and retrieves the medical image and the hidden data. The compressed image obtained from the previous phase is reconstructed to get the stego image by performing the run length decoding procedure as given below:
Steps

- Using the elements and run length vectors, create a vector of stego image.
- Convert the vector obtained in the previous step to 256x256 matrix.
- Retrieve the value of p(x,y) which is the ASCII value of the hidden character, if the following conditions are true:
  $p(x,y) \neq 0$ and
  $p(x+1,y)= P(x-1,y)= P(x,y+1)=P(x,y-1)=0$
- Replace p(x.y) with zero to get the original medical image.
- Convert the ascii values to their corresponding characters to get the secret data.

## 3.4 Flow Chart

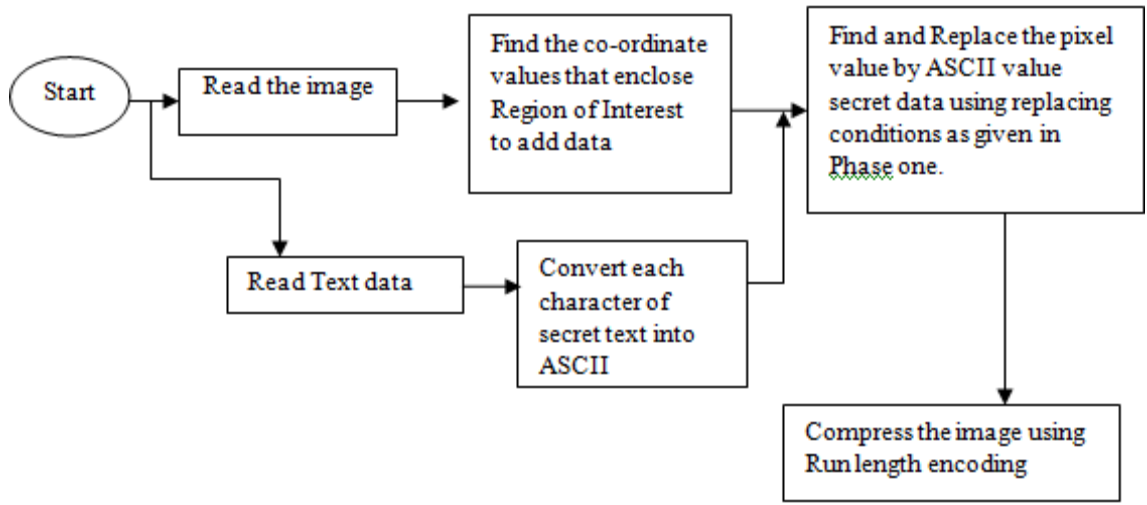

Figure 1.(a) Data hiding and compression

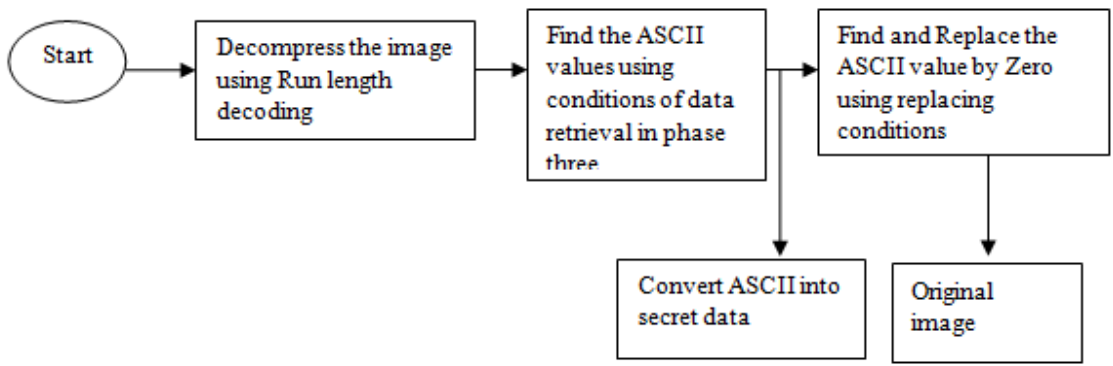

Figure 1. (b) Decompression and Data Retrieval

## 4. RESULTS AND DISCUSSION

The proposed method is tested using medical images of size 256x256 with 256 gray levels. Results of the proposed method is given in Table

Table 1: Secret Message

| Input Text | G | R | I |    | p | i | d | : | 0 | 0 | 7 |
|---|---|---|---|---|---|---|---|---|---|---|---|
| ASCII | 71 | 82 | 73 | 32 | 112 | 105 | 100 | 58 | 48 | 48 | 55 |

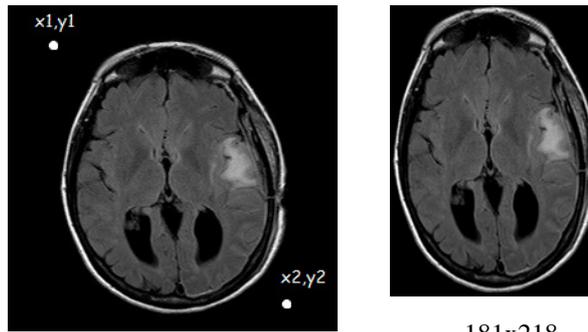

Figure 2. (a) Original medical image 256x256

181x218
Figure 2.(b) Image with Region of Interest 181x218

In the proposed method, even though an original medical image of size 256x256 is used, data is hidden near Region of Interest so that there will not be any visible distortion. This is illustrated in Figure 2 (b) which shows the actual region considered for hiding

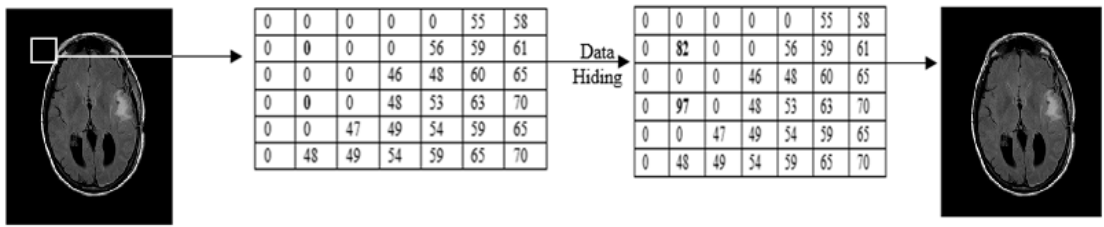

Figure 3. Image before and after hiding

| 109 | 109 | 99 | 99 | 99 | 99 | 99 | 97 | 97 | 97 |

(a) Image to vector

| 109 | 99 | 97 |

(b) Element

| 2 | 5 | 3 |

(c) Run Length

Figure 4: Sample result of Phase 2

The performance of the proposed system is analysed by measuring encoding and decoding time that are tabulated in Table 1. The proposed algorithm is very simple to execute and Table 1 reveals that the data hiding and encoding time takes less than a second.

Table 2 : Measurement of Processing Time

| Image | Process | Elapsed Time (seconds) | Total (seconds) |
|---|---|---|---|
| 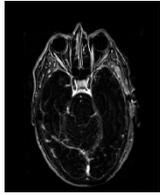 | Data hiding | 0.0172 | 0.1031 |
| | Run length Encoding | 0.0079 | |
| | Run length Decoding | 0.0720 | |
| | Data Retrieval | 0.0060 | |
| 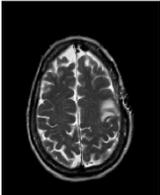 | Data hiding | 0.0169 | 0.0909 |
| | Run length Encoding | 0.0076 | |
| | Run length Decoding | 0.0607 | |
| | Data Retrieval | 0.0057 | |
| 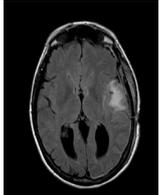 | Data hiding | 0.0175 | 0.1023 |
| | Run length Encoding | 0.0087 | |
| | Run length Decoding | 0.0688 | |
| | Data Retrieval | 0.0073 | |

The quality of the reconstructed image is assessed by measuring the standard metric Mean Square Error (MSE) between the original image and stego image as well as original and reconstructed image which is calculated as follows,

$$MSE = \frac{1}{mn} \sum_{i=0}^{m-1} \sum_{j=0}^{n-1} [I(i,j) - K(i,j)]^2$$

$$PSNR = 10 \cdot \log_{10}\left(\frac{MAX_I^2}{MSE}\right)$$

$$= 20 \cdot \log_{10}\left(\frac{MAX_I}{\sqrt{MSE}}\right)$$

$$= 20 \cdot \log_{10}(MAX_I) - 10 \cdot \log_{10}(MSE)$$

The MSE and PSNR values computed for stego images and reconstructed images is given in Table 3.

Table 3: MSE and PSNR of Stego Image and Reconstructed Image

| Compared Medical Images | | MSE | PSNR | Compared Medical Images | | MSE | PSNR |
|---|---|---|---|---|---|---|---|
| Original image | stego image | 0.9565 | 48.3240 | Original image | Reconstructed image | 0 | Infinity |
| 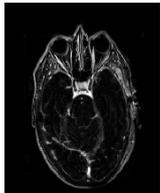 | 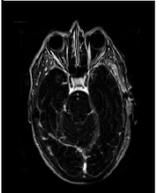 | | | 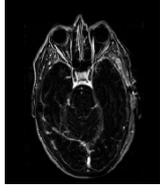 | 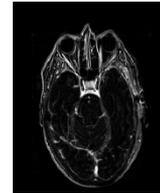 | | |
| 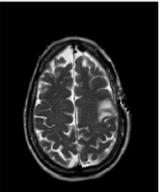 | 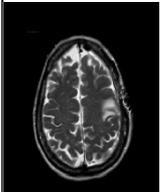 | 0.9565 | 48.3240 | 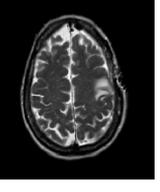 | 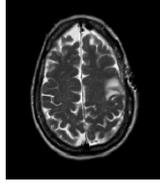 | 0 | Infinity |
| 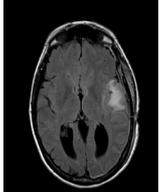 | 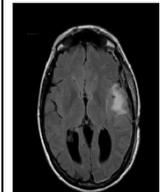 | 0.9565 | 48.3240 | 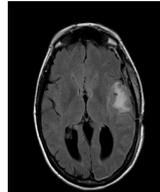 | 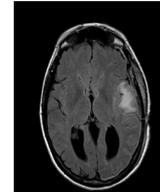 | 0 | Infinity |

The results of Table 3 shows that there is not much visible distortion between the stego images and the reconstructed images. It is also worth noting that MSE is 0 revealing that the reconstructed image is exactly similar to the original image. The retrieved data is same as the hidden data in these three sample images. The MSE and PSNR values in Table 2 for all sample images is same (PSE=0.9565, MSE=48.4240) indicating that the same number of pixels are modified in these three images.

Table 4 compares the proposed method with the existing method [7] and the result shows the efficiency of the proposed method.

Table 4: Comparison with Existing method

| S.No | Method | | Elapsed Time (Seconds) | | Total (Seconds) |
|---|---|---|---|---|---|
| 1 | Existing [7] | | Stego | 6.79 | 7.2100 |
| | | | Compression | 0.42 | |
| 2 | Proposed | Image 1 | Stego | 0.0251 | 0.1031 |
| | | | Compression | 0.0780 | |
| | | Image 2 | Stego | 0.0245 | 0.0909 |
| | | | Compression | 0.0664 | |
| | | Image 3 | Stego | 0.0262 | 0.1023 |
| | | | Compression | 0.0761 | |

## 5. CONCLUSION

This paper proposes a new approach that couples data hiding and compression in medical images. Experiments prove the superior performance of proposed method with low computation complexity and ensures fast and secure transmission of medical images. As Run Length coding lossless compression is used, the retrieved image by the receiver is the same as the original image sent by the sender. Gray scale images are used for testing the proposed method and our future work would be to extend to color images as well.

## AUTHORS

**Dr.M. Mary Shanthi Rani,** a NET qualified Assistant Professor in the Department of Computer Science and Applications, Gandhigram Rural Institute (Deemed University),Gandhigram has twelve years of teaching and eight years of research experience as well. She has nearly twenty publications in International Journals and Conferences. Her research areas of interest are Image Compression, Information

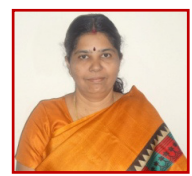

Security, Ontology, Biometrics and Computational Biology. She has authored a book titled "Novel Image Compression Methods Based on Vector Quantization" and is one of the editors of Conference Proceedings "Recent Advances in Computer Science and Applications". She has served in various academic committees in designing the curriculum for B.Sc. and M.C.A courses as well. She has also served as reviewer of Peer-reviewed International Journals and Conferences and is a Life member of Indian Society for Technical Education. She has the credit of being the Associate Project Director of UGC Indo-US 21$^{st}$ Knowledge Initiative Project.

**S.Lakshmanan** is a Research Scholar (Full-time) in the Department of Computer Science and Applications, Gandhigram Rural Institute - Deemed University, Dindigul, India. He received his Bachelor of Science (B.Sc) degree in Computer Science in the year 2012 from Madurai Kamaraj University and Master of Computer Applications (MCA) degree in the year 2015 from Gandhigram Rural Institute - Deemed University. He is currently pursuing Ph.D. degree in Gandhigram Rural Institute- Deemed University. His research focuses on Medical Image Processing. 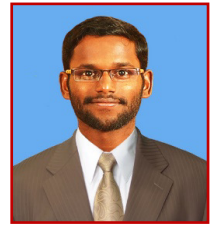